\documentclass[aip,graphicx]{revtex4-1}
\usepackage{comment}
\usepackage[inline, shortlabels]{enumitem}
\usepackage{graphicx}
\usepackage{dcolumn}
\usepackage{bm}
\usepackage{xcolor}
\usepackage{amsmath}

\draft 

\begin{document}


\title{An investigation of high entropy alloy conductivity using first-principles calculations} 



\author{Vishnu Raghuraman}
\email[Author to whom correspondence should be addressed: ]{vishnura@andrew.cmu.edu}
\affiliation{Department of Physics, Carnegie Mellon University, Pittsburgh, PA, 15213, USA}

\author{Yang Wang}
\affiliation{Pittsburgh Supercomputing Center, Carnegie Mellon University, Pittsburgh, PA, 15213, USA}
\author{Michael Widom}
\affiliation{Department of Physics, Carnegie Mellon University, Pittsburgh, PA, 15213, USA}

\date{\today}

\begin{abstract}
The Kubo-Greenwood equation, in combination with the first-principles Korringa-Kohn-Rostoker Coherent Potential Approximation (KKR-CPA) can be used to calculate the DC residual resistivity of random alloys at T = 0 K. We implemented this method in a multiple scattering theory based ab initio package, MuST, and applied it to the ab initio study of the residual resistivity of the high entropy alloy Al$_x$CoCrFeNi as a function of $x$. The calculated resistivities are compared with experimental data. We also predict the residual resistivity of refractory high entropy alloy MoNbTaV$_x$W.  The calculated resistivity trends are also explained using theoretical arguments. 

\end{abstract}


\maketitle 

High entropy alloys (HEAs) are alloys with five or more principal elements \cite{yeh,cantor,miracle}. The presence of multiple elements results in a large entropic contribution to the free energy, which stabilizes a single phase solid solution over other potentially competing intermetallic compounds. Since their introduction in 2004\cite{yeh,cantor}, they have become a highly active area of both experimental and theoretical research. This is largely due to the desirable functional and mechanical properties that HEAs may possess. For example, the quinary CoCrFeMnNi alloy (also called the Cantor alloy\cite{cantor}) is found to exhibit ultimate strengths and elongations in excess of 1 GPa and 60\% respectively at 77 K \cite{george}, as well as large fracture toughness \cite{fracture}; Cu$_x$CoFeMnNi has been used as a substrate to grow graphene \cite{graphene}; The electrical resistivity of the refractory HEA Hf$_8$Nb$_{33}$Ta$_{34}$Ti$_{11}$Zr$_{14}$ sharply drops to zero at $T_c \approx 7.3$ K \cite{superconductor}, demonstrating superconductivity. HEAs are promising candidates for hydrogen embrittlement resistance \cite{hydr-embrittle-res-1,hydr-embrittle-res-2}, magnetoresistance, shape memory response \cite{shape-memory}, response to irradiation \cite{irr-1,irr-2}, and other useful, interesting functional properties. 

Due to the large number of possible HEAs, first-principles calculations are an effective method of screening for desired properties. The \textit{ab-initio} approach has been used to obtain the density of states, evaluate formation energies, identify the different competing phases and transition temperatures, determine chemical species ordering, calculate elastic constants \textit{etc}. Electrical conductivity is an important functional property of alloys which is of both theoretical and practical interest. Predicting the conductivity from first-principles can be done in a few different ways. The problem can be treated semi-classically by solving the linearized Boltzmann equation \cite{ashcroft-mermin} numerically, with the required band-structures obtained from DFT calculations. This approach was first used by Stocks and Butler\cite{AgPd} to calculate the residual resistivity of Ag$_x$Pd$_{1 - x}$ system using the single-site Korringa-Kohn-Rostoker Coherent Potential Approximation\cite{kkr-cpa-1,kkr-cpa-2,kkr-cpa-3} (KKR-CPA).  Their results agreed well with experiment and since then, this technique has been used on wide variety of alloy systems. Recently, Wang \textit{et al} \cite{mea} used the software \texttt{BoltzTraP2} \cite{BoltzTrap} to calculate the electrical transport properties of medium entropy alloy family Si$_y$Ge$_y$Sn$_x$. In this case, the band structure was obtained from supercell calculations using a plane-wave pseuopotential code. The semi-classical method, however, requires a well-defined band structure at the Fermi energy. As a result, it cannot be applied to alloys with high chemical disorder, as such systems lack a sharp Fermi surface and their electronic states have finite lifetimes, hence their energy levels are broadened. 

To avoid this issue, the conductivity can also be obtained using Kubo linear response theory \cite{kubo}. The Kubo formalism deals with current-current correlation functions instead of band structures, and is capable of dealing with highly disordered alloys. In a seminal paper \cite{butler}, Butler combined the Kubo-Greenwood equation with KKR-CPA and derived conductivity expressions for random alloys. This technique includes vertex corrections, which represent the ``scattering-in" contribution to the conductivity. It was first implemented by Swihart et al \cite{swihart}, and employed to calculate residual resistivities for Cu$_x$Zn$_{1 - x}$, Cu$_x$Ga$_{1 - x}$, Cu$_x$Ge$_{1 - x}$ and Ag$_{x}$Pd$_{1 - x}$ alloys. Some HEA conductivity studies have also been done using this approach. Mu et al \cite{saimu} used the Kubo-Greenwood formalism to study the residual resistivity of the Cantor-Wu alloys \cite{cantorwu}. In this paper, we present an implementation of the Kubo-Greenwood formalism in the open source, multiple scattering theory based, DFT code \texttt{MuST} \cite{MuST}, and we show the residual resistivity of the multiphase high entropy alloy  Al$_x$CoCrFeNi as a function of $x$. The electrical and thermal properties of Al$_x$CoCrFeNi make it a suitable candidate for thermoelectric applications like waste heat recovery and refrigeration\cite{AlCoCrFeNi-2}. We are able to recreate the experimental trend of increased resistivity at larger values of $x$. In addition, our calculations recover the non-monotonic behaviour observed in the multi-phase region. We also obtain first-principles predictions for the single-phase refractory high entropy alloy MoNbTaV$_x$W as a function of $x$. For both these systems, we provide theoretical justifications for the calculated results.

The paper is organized as follows. First, we provide a brief introduction to the theory behind the KKR method, the Coherent Potential Approximation, and the Kubo-Greenwood equation in KKR-CPA formalism. Our implementation of this technique is first tested on two non-spin-polarized binaries - BCC CuZn and FCC AgPd. It is further tested on Cantor-Wu alloys, which are magnetic FCC solid solutions. After establishing confidence in the code,  we apply it to the the two HEA systems previously mentioned, along with a comparison to previously obtained experimental data for the Al alloy. Finally, we conclude by commenting on the limitations of our current implementation and discuss how these limitations can be overcome, so that this code can be used to study more complex functional alloy behavior.

The density functional theory (DFT) based ab initio method is built upon solving a single electron Schr\"odinger equation, called the Kohn-Sham (KS) equation, \cite{dft-1,dft-2}
\begin{equation}
    \left[-\nabla^{2} + V_{\rm{eff}}(\left[\rho(\bm{r})\right])\right]\psi_{i}(\bm{r}) = \epsilon_{i}\psi_{i}(\bm{r}),
    \label{eq:ks}
\end{equation}
where the electron density $\rho(\bm{r})$ is given by
\begin{equation}
    \rho(\bm{r}) = \sum_{\epsilon_i \leq \epsilon_F} \vert \psi_i(\bm{r}) \vert^2,
    \label{eq:density}
\end{equation}
assuming that the KS wavefunctions are orthogonal and the Fermi energy is determined by the number of electrons in the system. The effective potential $V_{\rm{eff}}$ consists of electrostatic terms and the exchange-correlation (XC) term. The XC functional can be modelled in different ways; in this paper, we use the Local Density Approximation (LDA) \cite{dft-1}, where the XC functional is purely dependent on the local density $\rho(\bm{r})$. A popular alternative is the Generalized Gradient Approximation (GGA) \cite{gga-1,gga-2}, which depends on the local density and it's gradient.

The conventional approach for solving (\ref{eq:ks}) is to diagonalize the Hamiltonian of the Kohn-Sham equation and calculate the corresponding eigenvalues and eigenvectors. From the eigenvectors, the density is calculated self-consistently using (\ref{eq:density}). Many popular DFT codes like \texttt{VASP, WIEN2k, Quantum Espresso} etc use this approach. However, the density can also be obtained from the Green's Function of the Hamiltonian - an approach taken by the KKR-Green's Function method~\cite{faulkner_stocks_1980,faulkner_stocks_wang}, based on multiple scattering  theory~\cite{korringa_1947,kohn_rostoker_1954}.

In the multiple scattering theory approach to an alloy, the system is divided into non-overlapping atomic cells, each of which has one atom present at the center. The total effective potential $V_{\rm{eff}}(\bm{r})$ can then be considered as the sum of the cell potentials 
\begin{equation}
    v_n(\bm{r_n}) = 
    \begin{cases}
    V_{\rm{eff}}(\bm{r}),\;\text{if}\;\bm{r} \in \Omega_{n} \\
    0, \; \text{otherwise}
    \end{cases},
    \label{eq:vn}
\end{equation}
with $\bm{r}_n = \bm{r} - \bm{R}_n$, where $\bm{R}_n$ is the position vector of the atomic site and $\Omega_n$ is the volume of the $n$th cell. Each cell can be treated as an electron scatterer, with $\underline{t}^n(\epsilon)$ as the $t$-matrix associated with the local potential in the $n$th cell. In addition, we can define the multiple scattering path matrix \cite{tau-def}
\begin{equation}
    \underline{\tau}^{nm}(\epsilon) = \underline{t}^{n}(\epsilon)\delta_{nm} + \underline{t}^{n}(\epsilon)\sum_{k\neq n} \underline{g}^{nk}(\epsilon) \underline{\tau}^{km}(\epsilon),
    \label{eq:tautrecur}
\end{equation}
as the sum of all the scattering processes that start from cell $n$ and end at cell $m$, and $\underline{g}^{nk}(\epsilon)$ is the free electron propagator from cell $n$ to cell $k$. The multiple scattering path matrix can be used to construct the single-site Green's Function at cell $n$ \cite{faulkner_stocks_1980,faulkner_stocks_wang}
\begin{eqnarray}
    G(\bm{r_n}, \bm{r_n}, \epsilon) & = & \sum_{LL^{\prime}} Z^{n}_L (\bm{r_n}, \epsilon)\tau^{nn}_{LL^{\prime}}(\epsilon)Z^{n\bullet}_{L^{\prime}} (\bm{r_n}, \epsilon) \nonumber \\
& &   - \sum_{L} Z^{n}_L (\bm{r_n}, \epsilon)J^{n\bullet}_L(\bm{r_n}, \epsilon),\label{eq:gf}
\end{eqnarray}
where $L$ is a combination of the angular momentum quantum numbers $l$ and magnetic quantum number $m$, $Z^{n}_L(\bm{r_n}, \epsilon)$ ($J^{n}_{L}(\bm{r_n}, \epsilon)$) represents a regular (irregular) solution to the single site Schr\"odinger's equation with potential $v_n(\bm{r_n})$ shown in equation (\ref{eq:vn}) at site $n$. The dot ($\bullet$) operator is applied to the spherical harmonics of the single-site solutions when they are written as a series expansion in the angular momentum basis. 

The electron density in cell $n$ can be obtained using
\begin{equation}
    \rho(\bm{r_n}) = -\frac{1}{\pi}\mathrm{Im}\int_{-\infty}^{\epsilon_{F}} G(\bm{r_n}, \bm{r_n}, \epsilon)\;d\epsilon,\;\; {\rm with}\;  \bm{r_n}\in\Omega_n,
\end{equation}
where $\epsilon_F$ is the Fermi energy. This density can then be used to re-calculate the effective potential $V_{\rm{eff}}(\left[\rho(\bm{r})\right])$ in the Hamiltonian, and the re-calculated Hamiltonian produces a new Green's Function. This cycle can be continued till self-consistency is reached. As mentioned in the previous section, the calculation of the electron density from the Green's function implies that the energy eigenvalues and the KS wavefunctions are not required. This approach has some significant advantages, one of which is the ability to combine with coherent potential approximation to deal with random systems.

The KKR-Green's function method combined with the CPA forms a powerful technique, namely KKR-CPA, to deal with random alloys from the first principles. It is based on constructing an effective medium which mimics the ensemble average of a disordered system. The CPA medium can be imagined as a periodic system consisting of a ``virtual" species, described by $t$-matrix $\underline{t}_{\rm{CPA}}(\epsilon)$. The CPA medium can be conveniently obtained by applying the single-site approximation
\begin{equation}
    \underline{\tau}_{\rm{CPA}}^{nn}(\epsilon) = \sum_{\alpha} c_{\alpha} \underline{\tau}^{nn}_{\alpha}(\epsilon),
    \label{eq:single-site}
\end{equation}
which assumes that the chemical species distribution on the underlying lattice is completely random. The probability of a particular site occupied by species $\alpha$ is determined by its concentration $c_{\alpha}$ in the alloy. The term $\underline{\tau}^{nn}_{\alpha}(\epsilon)$, which represents the multiple scattering path matrix for a CPA medium with an impurity of species $\alpha$ at site $n$, is obtained from \cite{ecm-cpa}
\begin{equation}
    \underline{\tau}^{nn}_{\alpha}(\epsilon) = \left[1 + \underline{\tau}^{nn}_{\rm{CPA}}(\epsilon)(\underline{t}^{-1}_{\alpha}(\epsilon) - \underline{t}^{-1}_{\rm{CPA}}(\epsilon))\right]^{-1}\underline{\tau}^{nn}_{\rm{CPA}}(\epsilon).
    \label{eq:single-site-impurity}
\end{equation}
Furthermore, due to the periodicity of the CPA medium, the $t$-matrix and the multiple scattering path matrix for the medium are related through the equation
\begin{equation}
    \underline{\tau}^{nn}_{\rm{CPA}}(\epsilon) = \frac{1}{\Omega_{BZ}}\int d^{3}\bm{k}\;\left[\underline{t}^{-1}_{\rm{CPA}}(\epsilon) - \underline{g}(\bm{k}, \epsilon)\right]^{-1},
    \label{eq:ft}
\end{equation}
where $\underline{g}(\bm{k}, \epsilon)$ is the lattice Fourier transform of the free electron propagator $\underline{g}^{nk}(\epsilon)$ in equation (\ref{eq:tautrecur}) and $\Omega_{BZ}$ is the volume of the first Brillouin zone. Equations (\ref{eq:single-site})-(\ref{eq:ft}) can be combined to create an iterative scheme that determines $\underline{t}_{\rm{CPA}}(\epsilon)$ self-consistently.

CPA is a successful and popular technique that has been used heavily to calculate total energy, density of states and other important electronic structure properties of random alloys. However, the single-site approximation ignores any short-range order that may be present in the alloy. While there are techniques available that incorporate chemical short range order in CPA \cite{pre-ecm-cpa,ecm-cpa,nlcpa,nlcpa-2,cacpa}, they are beyond the scope of this work.

For a system of non-interacting electrons moving under the influence of a random potential, the symmetric part of the DC electrical conductivity tensor at T = 0 K can be expressed as\cite{kubo}
\begin{equation}
    \sigma_{\mu\nu}(\epsilon) = \frac{\pi}{N\Omega}\left\langle \sum_{\lambda,\lambda^{\prime}} \langle \lambda \vert \hat{j}_{\mu} \vert \lambda^{\prime} \rangle \langle \lambda^{\prime} \vert \hat{j}_{\nu} \vert \lambda\rangle \delta(\epsilon - \epsilon_{\lambda})\delta(\epsilon - \epsilon_{\lambda^{\prime}})\right\rangle,
    \label{eq:kg}
\end{equation}
where $\mu$, $\nu$ represent Cartesian directions, $\vert \lambda \rangle, \vert \lambda^{\prime} \rangle$ are the eigenvectors of the Hamiltonian associated with a given configuration of the disordered system, $\epsilon_{\lambda}, \epsilon_{\lambda^{\prime}}$ are the corresponding eigenvalues, $N$ represents the number of atoms in the system and $\Omega$ is the volume per atom. The angle brackets denote an average over all the possible configurations of the disordered system, and $\hat{j}_{\mu}$ refers to the current operator component, which in the non-relativistic case is given in atomic units as 
\begin{equation}
    \hat{j}_{\mu} = -2\sqrt{2}i\frac{\partial}{\partial r_{\mu}}
\end{equation}
In order to use (\ref{eq:kg}) within the framework of multiple scattering theory, the eigenkets must be replaced with multiple scattering path matrices. With some algebra, and the use of (\ref{eq:gf}), we can express (\ref{eq:kg}) as \cite{butler}
\begin{align}
    \sigma_{\mu\nu} = \lim_{\delta \xrightarrow{} 0}\frac{1}{4}\left[\tilde{\sigma}_{\mu\nu}(\epsilon^{+}, \epsilon^{+}) - \tilde{\sigma}_{\mu\nu}(\epsilon^{+}, \epsilon^{-})
    - \tilde{\sigma}_{\mu\nu}(\epsilon^{-}, \epsilon^{+}) + \tilde{\sigma}_{\mu\nu}(\epsilon^{-}, \epsilon^{-})\right],
\end{align}
where
\begin{align}
    \tilde{\sigma}_{\mu\nu}(z_1,z_2) = 
    -\frac{1}{\pi N\Omega}\langle J^{m\mu}_{L_4L_1}(z_2, z_1)\tau^{mn}_{L_1L_2}(z_1)J^{n\nu}_{L_2L_3}(z_1, z_2)\tau^{nm}_{L_3L_4}(z_2)\rangle.
    \label{eq:kgg}
\end{align}
with $\epsilon^{+} = \epsilon_{F} + i\delta$ and $\epsilon^{-} = \epsilon_{F} - i\delta$. Here the symbol $J$ refers to the current matrix, which is the position basis representation of the current operator matrix elements. Butler further demonstrates that (\ref{eq:kgg}) can be rephrased within the single-site CPA formalism as\cite{butler}
\begin{align}
    \tilde{\sigma}_{\mu\nu} = -\frac{1}{\pi\Omega}&\left(\sum_{\alpha\beta}c_{\alpha}c_{\beta}\tilde{J}^{\alpha\mu}_{K_1}(z_2, z_1)\left[1 - \chi\omega\right]^{-1}_{K_1K_2}\tilde{J}^{\beta\nu}_{K_2}(z_1, z_2) \right. \nonumber \\
    &+ \left.\sum_{\alpha}c_{\alpha}\tilde{J}^{\alpha\mu}_{L_4L_1}(z_2, z_1)\tau^{\mathrm{CPA}}_{L_1L_2}(z_1)J^{\alpha\nu}_{L_2L_3}(z_1, z_2)\tau^{\mathrm{CPA}}_{L_3L_4}(z_2) \right).
\end{align}
The terms $\chi$ and $\omega$ are 4th order tensors given by the expressions \cite{butler}
\begin{align}
    \chi_{L_1L_2L_3L_4} &= \frac{1}{\Omega_{BZ}}\int d^{3}\bm{k}\;\tau^{\rm{CPA}}_{L_1L_2}(\bm{k}, z_1)\tau^{\rm{CPA}}_{L_3L_4}(\bm{k}, z_2)
    - \tau^{\rm{CPA}}_{L_1L_2}(z_1)\tau^{\rm{CPA}}_{L_3L_4}(z_2), \\
    \omega_{L_1L_2L_3L_4} &= \sum_{\alpha} c_{\alpha} x^{\alpha}_{L_1L_2}(z_1)x^{\alpha}_{L_3L_4}(z_2).
\end{align}
The term $(1 - \chi\omega)^{-1}$ represents the vertex correction, or ``scattering-in" term. To understand these equations and how they are derived in more detail, the reader is referred to Butler's work~\cite{butler}. Additionally, to learn more about the calculation of the current matrices and other important details pertaining to implementation, the reader is referred to the work by Banhart~\cite{banhart}. It can be seen that calculating $\chi$ is computationally expensive, due to the Brillouin Zone integration. However, this calculation can be made significantly faster by only integrating over the irreducible part of the Brillouin Zone, and then applying rotation operations to recover the contribution of the other parts. For crystals with cubic symmetry, the integration only has to be performed over 1/48th of the first Brillouin zone.  For certain systems, approximating $(1 - \chi\omega)^{-1} \approx I$ does not significantly impact the conductivity. An example of this is the Ag-Pd binary at high Pd concentration, the reasons for which have been explored in details elsewhere~\cite{agpd-vertex-corrections}. Hence, computational cost can be reduced in such calculations by neglecting the vertex corrections. In this paper, however, the vertex corrections are included in all the conductivity calculations.

The conductivity expressions have been implemented in MuST~\cite{MuST}, a multiple scattering theory based open source code for ab initio electronic structure calculations.
In order to demonstrate the validity of the implementation, the code is applied to systems for which computational results are available for validation. First, we apply them to the BCC Cu-Zn binary and FCC Ag-Pd binaries.
\begin{figure}
    \centering
    \includegraphics{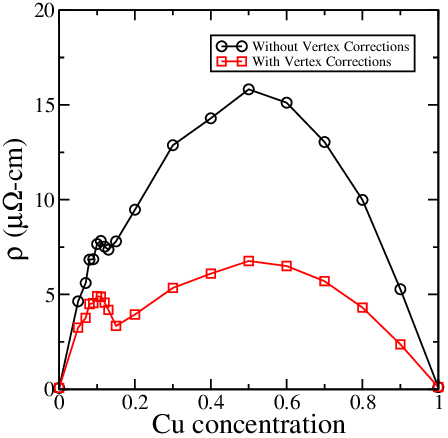}
    \caption{Residual resistivity of Cu$_x$Zn$_{1 - x}$ from first-principles calculations. The red curve (with circles) represents resistivity without vertex corrections and the black curve (with squares) represents resistivity with vertex corrections. Note that calculations have been performed for the BCC phase at all values of $x$.}
    \label{fig:CuZn-conductivity}
\end{figure}
Figure \ref{fig:CuZn-conductivity} shows the residual resistivity of Cu-Zn obtained using MuST. Despite CuZn showing BCC phase only for the equiatomic case, the calculations have been done for the BCC phase at all Cu concentrations. The resistivity appears to obey Nordheim's relation ($\rho_{0} \propto x(1 - x)$) \cite{nordheim}, and compares well with previously obtained computational \cite{cuzn-computational} and experimental \cite{cuzn-experimental} results. However, an unexpected peak in resistivity can be seen in the vicinity of 10\% Cu, which was not examined in the previous computational studies\cite{cuzn-computational,swihart}. Comparison with measurements is difficult for this concentration, as there is a scarcity of experimental resistivities for the Zn-rich region\cite{cuzn-experimental}. Further, the experimentally observed phase is not BCC at this composition. The peak is large enough to rule out numerical errors as the cause, implying that there is some interesting physics going on here which needs further study.  It can be seen that vertex corrections play a major role in this system, a feature that has also been previously noted \cite{swihart}.
\begin{figure}
    \centering
    \includegraphics{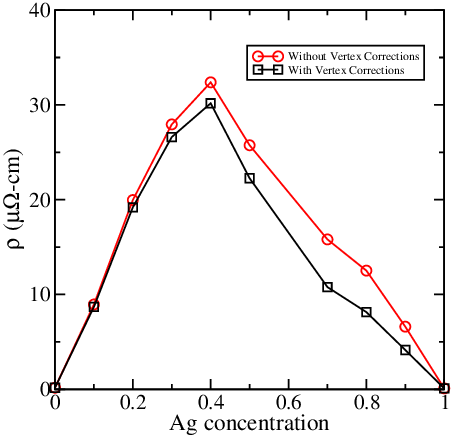}
    \caption{Residual resistivity of Ag$_x$Pd$_{1 - x}$ from first-principles calculations. The red curve (with circles) represents resistivity without vertex corrections and the black curve (with squares) represents resistivity with vertex corrections. Note that calculations have been performed for the FCC phase at all values of $x$.}
    \label{fig:AgPd-conductivity}
\end{figure}
Figure \ref{fig:AgPd-conductivity} shows the residual resistivity of Ag-Pd obtained using MuST, which compares well with previously obtained computational results \cite{cuzn-computational}. A significant deviation from Nordheim's relation can be observed. This is attributed to the palladium $d$-states, which are the major contributors to the DOS at Fermi energy for high palladium concentrations\cite{swihart}. It can also be seen that the vertex corrections are negligible at high Pd concentration, but become slightly more significant as the silver content increases. The absence of vertex corrections is also associated with a predominant $d$ character at the Fermi energy\cite{agpd-vertex-corrections}.
\begin{figure}
    \centering
    \includegraphics{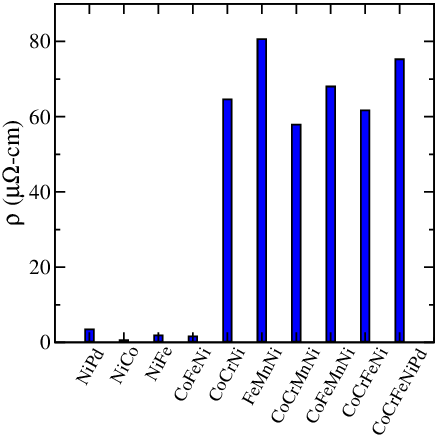}
    \caption{The residual resistivity of a set of Cantor-Wu alloys calculated using Kubo-Greenwood formula with KKR-CPA method implemented in MuST. All calculations are spin-polarized.}
    \label{fig:cantor-wu}
\end{figure}
To further test the the implementation of the Kubo-Greenwood formula in MuST, we apply it to the conductivity calculation for the Cantor-Wu alloys, which are magnetic solid solutions on an FCC lattice. Specifically, we examine the binaries NiFe, NiCo, and NiPd, the ternaries FeMnNi, CoCrNi and CoFeNi, the four element HEAs CoCrFeNi and CoFeMnNi, and the quinary CoCrFeNiPd. Figure \ref{fig:cantor-wu} shows the residual resistivity for these systems. These values are nearly similar to the computational results obtained by Mu et al~\cite{saimu}. The alloys fall into two categories: low residual resistivity alloys containing Co, Fe, Ni or Pd and high residual resistivity alloys containing Cr and/or Mn. This curious trend was explained by looking at the spin-resolved Fermi surfaces \cite{saimu}. The low resistivity alloys have sharp Fermi surfaces, which are associated with large mean free paths; conversely, the high resistivity alloys have washed out Fermi surfaces, which are associated with shorter mean free paths. The CuZn, AgPd and Cantor-Wu results instill confidence in the implementation of our code, and demonstrate its viability as a useful tool to study HEA electrical conductivity.

The first HEA we study is the five-element Al$_x$CoCrFeNi, with $x$ denoting the Al content. Experimentally, this alloy is single-phase FCC for $0 \leq x \leq 0.375$ and single-phase BCC for $1.25 \leq x \leq 2$. At intermediate values of $x$, the alloy shows multi-phase FCC+BCC behavior \cite{AlCoCrFeNi-1,AlCoCrFeNi-2}. Multi-phase systems are difficult to study, owing to their inhomogeneity. As a result simple models are needed to calculating the resistivity in this regime. We calculate the pure FCC resistivity $\rho_{\rm{FCC}}$ and pure BCC resistivity $\rho_{\rm{BCC}}$ and perform a parallel average
\begin{equation}
    \frac{1}{\rho_{p}} = \frac{w_{\rm{FCC}}}{\rho_{\rm{FCC}}} + \frac{w_{\rm{BCC}}}{\rho_{\rm{BCC}}}
\end{equation}
and a series average
\begin{equation}
    \rho_s = w_{\rm{FCC}}\rho_{\rm{FCC}} + w_{\rm{BCC}}\rho_{\rm{BCC}},
\end{equation}
where $w_{\rm{FCC}}$ and $w_{\rm{BCC}}$ are the averaging weights that denote the volume fractions of the FCC/BCC phases in the alloy. These parameters are taken from experiment\cite{concentrations}.
\begin{figure}
    \centering
    \includegraphics{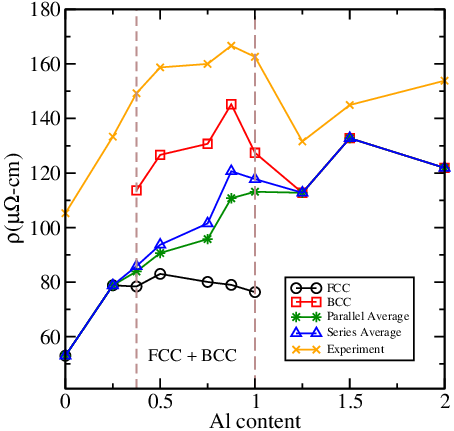}
    \caption{Residual resistivity of Al$_x$CoCrFeNi as a function of $x$. The black curve with circles and the red curve with squares represent the residual resistivity of the pure FCC and BCC phase respectively, obtained from first-principles calculations. The blue curve with triangles and green curve with stars represents the parallel and series averages. The orange curve with crosses is the experimental result\cite{AlCoCrFeNi-1}.}
    \label{fig:AlCoCrFeNi-conductivity}
\end{figure}
Figure \ref{fig:AlCoCrFeNi-conductivity} shows the residual resistivity calculated using MuST, where LDA was used as the exchange-correlation functional. The resistivities calculated using PBE significantly differed from the LDA values, and did not recreate the experimental trend at some Al concentrations. The sensitivity of the residual resistivity values to the exchange-correlation functional and the lattice parameter is discussed in the supplementary material. Both experimental and first-principles resistivities increase with increasing Al content. The BCC resistivity values closely follow the non-monotonic behaviour experimentally observed in the multi-phase region. While both the series and parallel averages are nearly similar, it can be seen that the shape of the parallel averaged curve is closer to experiment. The experimental values are larger than the calculated resistivities. This is due to two reasons. Firstly, the complexity of the experimental microstructures are not captured by the first principles calculations. This leads to lower scattering in the computational systems, leading to lower resistivities. Secondly, the experimental curve is obtained at room temperature, and it is expected to exceed the calculated residual resistivity.  To capture the non-monotonicity, the experimentally observed concentrations of the consistent elements were used \cite{concentrations}. If nominal concentrations are used, the non-monotonic behaviour is lost. This is most likely the reason for monotonic results obtained in a previous computational study of this HEA \cite{monotonic}.  

\begin{figure}
    \centering
    \includegraphics{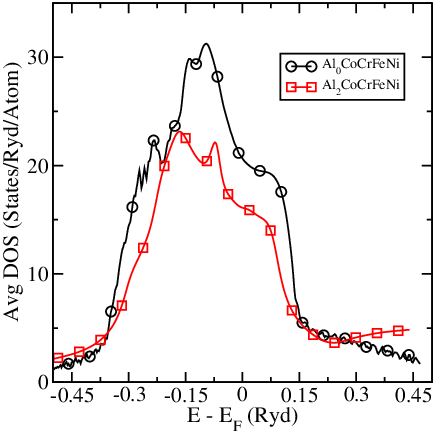}
    \caption{The averaged density of states for CoCrFeNi and Al$_2$CoCrFeNi. This was obtained by taking the concentration-weighted average of the partial density of states for the two systems.}
    \label{fig:AlCoCrFeNi-DOS-comparison}
\end{figure}

Figure \ref{fig:AlCoCrFeNi-DOS-comparison} shows the concentration averaged density of states (DOS) for two different Al concentrations. The DOS at the Fermi level is much lower for Al concentration $x = 2$ than $x = 0$, owing to the lack of $d$-electrons in Al. 
\begin{figure}
    \centering
    \includegraphics{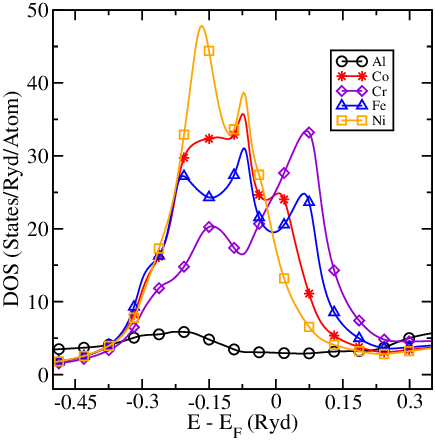}
    \caption{Partial DOS for the elements present in Al$_2$CoCrFeNi.}
    \label{fig:Al2_dos}
\end{figure}
The reason for the lower Fermi level DOS also becomes apparent when we look at the partial DOS for Al$_2$CoCrFeNi in Figure \ref{fig:Al2_dos}. The DOS at Fermi level for Al is significantly lower than the other transition elements. This, however, does not guarantee increased resistivities at higher Al concentrations. The effective valence of transition metal atoms in Al-TM alloys has been predicted and observed to be negative \cite{raynor}. This is due to the strong \textit{sp-d} hybridization, which leads to the transfer of $sp$ electrons (conduction electrons) to the partially filled $d$ bands of transition elements\cite{trambly,moriarty}. This explains the reduction in conductivity, or increase in residual resistivity with increasing Al content.

Finally, we use our first-principles code to predict the residual resistivity trend of MoNbTaV$_x$W as a function of $x$. This five element refractory high entropy alloy is a single-phase BCC solid solution.
\begin{figure}
    \centering
    \includegraphics{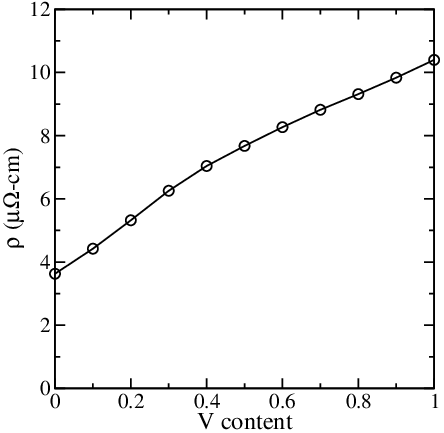}
    \caption{First-principles residual resistivity for MoNbTaV$_x$W as a function of $x$. }
    \label{fig:MoNbTaVW-conductivity-plot}
\end{figure}
\begin{figure}
    \centering
    \includegraphics{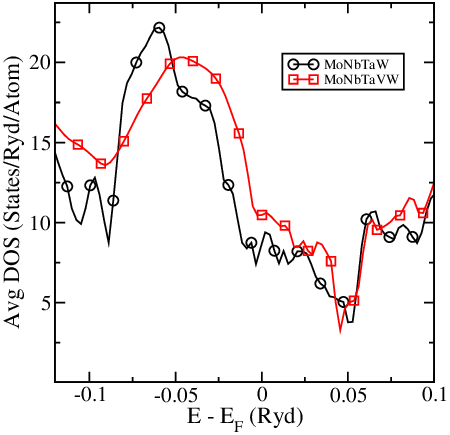}
    \caption{Concentration averaged DOS for MoNbTaW (black curve with circles) compared with equiatomic MoNbTaVW (red curve with squares)}
    \label{fig:MoNbTaVW-DOS-comparison}
\end{figure}
\begin{figure}
    \centering
    \includegraphics{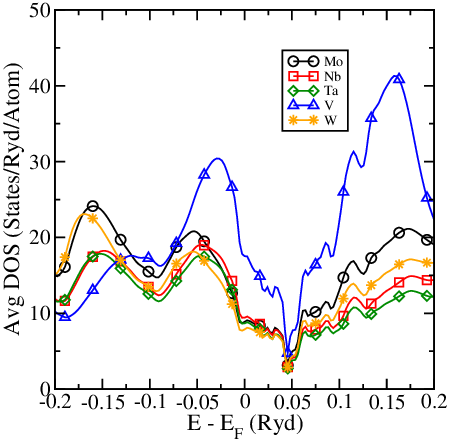}
    \caption{Partial DOS for the elements present in equiatomic MoNbTaVW alloy.}
    \label{fig:MoNbTaVW-DOS}
\end{figure}
Figure \ref{fig:MoNbTaVW-conductivity-plot} shows the calculated resistivities, and Figures \ref{fig:MoNbTaVW-DOS-comparison} and \ref{fig:MoNbTaVW-DOS} show the concentration averaged DOS and partial DOS respectively. In Figure \ref{fig:MoNbTaVW-conductivity-plot}, LDA has been used as the exchange-correlation functional. For this system, PBE produces a larger resistivity at all vanadium concentrations. This is partly due to well known tendency of LDA to produce ``overbinding" results\cite{overbinding}, but other effects may also contribute. The sensitivity of the residual resistivity values to the exchange-correlation functional and the lattice constant is discussed in the supplementary material. There is an apparent conflict here - the concentration averaged DOS for $x = 1$ is higher than for $x = 0$, owing to the larger vanadium partial DOS as compared to the other elements. However, the calculated residual resistivities increase with vanadium content. This can be resolved by considering the atomic sizes of the element. While the refractory metals have similar atomic sizes, vanadium is significantly smaller. This increases disorder in the system, which results in increased scattering, or increased residual resistivity. The size difference also means that vanadium is weakly bound to the refractory metals. When atoms come together in a solid, their sharp atomic energy spectra broaden and form energy bands. Stronger bonding will result in an energy spectrum which is more spread out. Conversely, weak bonding will result in more localized atomic-like energy bands, which are narrower and sharper. This explains both the higher resistivity and higher density of states. This alloy provides a important takeaway - an increase or decrease in the DOS does not automatically imply an increased or decreased residual resistivity. Additional analysis is necessary to explain the calculated trends.

In conclusion, we have implemented the Kubo-Greenwood equation in the non-relativistic single-site KKR-CPA formalism and tested it on the previously studied systems CuZn, AgPd, and the Cantor-Wu alloys. After code validation, we applied our code to obtain the residual resistivities of Al$_x$CoCrFeNi as a function of the concentration of Al. We found that both experiment and first-principles calculations show increased resistivities at higher values of $x$. This can be explained on the basis of the \textit{sp-d} hybridization that occurs in Al-TM alloys. In the multiphase region, however, the experimental resistivity is found decreasing with increasing the content of Al. This effect was captured in first-principles calculations, as a result of using experimentally obtained concentrations for the BCC and FCC phase. We also predicted the residual resistivity of refractory HEA MoNbTaV$_x$W as a function of $x$. We found that the residual resistivity increases with V content, which was explained on the basis of the size mismatch between V and the other refractory metals. Despite the reduction in conductivity, the DOS at the Fermi energy increased with the addition of V. This occurred due to the higher partial DOS for V, which results from it's weak bonding to the refractory metals. 

There are many possible extensions to our current code implementation. 
We intend to incorporate chemical short range order by combining the CA-CPA technique~\cite{cacpa} with the Kubo-Greenwood equation. This will allow us to study K-state alloys like CoCrNi and Ni25Cr, which are systems where residual resistivity increases with short range order.
\section*{Supplementary Material}
The supplementary material contains the computational details relevant to the first-principles calculations and the lattice parameters used for all the systems studied in this paper. The variation of the resistivity with the lattice parameter is examined, along with a comparison of computational results obtained using PBE and LDA functionals.

This work is based on open-source ab initio software package MuST \cite{MuST}, a project supported in part by NSF Office of Advanced Cyberinfrastructure and the Division of Materials Research within the NSF Directorate of Mathematical and Physical Sciences under award number 1931367, 1931445, and 1931525. The conductivity implementation in MuST and the calculations done in this paper were supported by the Department of Energy under Grant No. DE-SC0014506. We acknowledge helpful discussions with S Mu, G. Malcolm Stocks, H Ebert and J Banhart during code implementation. 

\section*{Data Availability Statement}
All the relevant data and computational details are present in the supplementary material.


%
%

%



\end{document}



\title{Supplementary Material : An investigation of high entropy alloy conductivity using first-principles calculations} 



\author{Vishnu Raghuraman}
\affiliation{Department of Physics, Carnegie Mellon University, Pittsburgh, PA, 15213, USA}

\author{Yang Wang}
\affiliation{Pittsburgh Supercomputing Center, Carnegie Mellon University, Pittsburgh, PA, 15213, USA}
\author{Michael Widom}
\affiliation{Department of Physics, Carnegie Mellon University, Pittsburgh, PA, 15213, USA}

\date{\today}
\maketitle
\section{Computational Details and Data}
All electrical conductivity calculations use atomic sphere approximation (ASA), LDA as the exchange-correlation potential and angular momentum cutoff $l_{\rm{max}} = 4$. Special K-points method was used for the Brillouin Zone integration. For CuZn, a 120x120x120 k-grid was used to calculate the conductivity, while for AgPd a 60x60x60 k-grid was employed. For the Cantor-Wu alloys and HEAs, a maximum of 70x70x70 K-points were used (although for some alloys lesser Kpoints were sufficient).  For the Cantor-Wu alloys, the experimental lattice constants were used and are present in the supplementary materials of Mu et al\cite{saimu}. For the Al$_x$CoCrFeNi, the experimental lattice constants were used, taken from Kao et al\cite{concentrations}. For AgPd, the lattice constants were kindly provided to us by J Banhart. For the other systems, lattice constants were calculated from first-principles and are presented in Tables S1 to \ref{tab:MoNbTaVW}.
\begin{figure}
    \centering
    \includegraphics{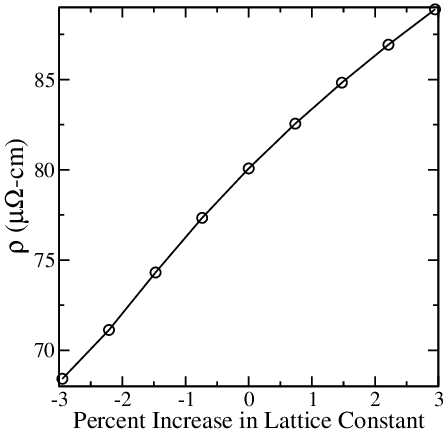}
    \caption{Residual resistivity of FCC Al$_{0.75}$CoCrFeNi as a function of the percentage change in the lattice constant (from the experimental value).}
    \label{fig:fcc-lc}
\end{figure}
\begin{figure}
    \centering
    \includegraphics{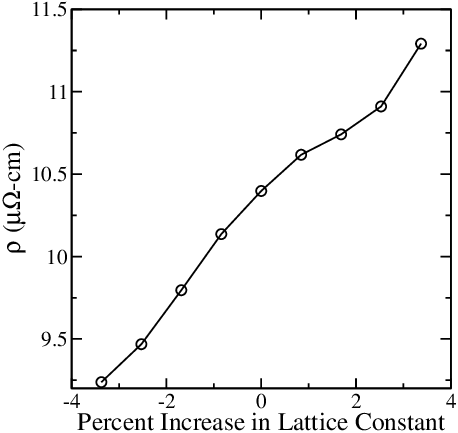}
    \caption{Residual resistivity of BCC MoNbTaVW as a function of percentage change in the lattice constant (from the first-principles equilibrium value).}
    \label{fig:bcc-lc}
\end{figure}
\section{Lattice Constant Sensitivity}
The effect of lattice constant on the resistivity of Al$_x$CoCrFeNi is expected to vary with $x$ and the type of crystal structure (FCC or BCC). However, it is instructive to look at a particular case - FCC Al$_{0.75}$CoCrFeNi. Figure \ref{fig:fcc-lc} shows the variation of residual resistivity with percentage change in the lattice constant. It is clear that the resistivity is highly sensitive to the lattice constant (with a change of almost 20 $\mu\Omega$-cm with 6\% change in the lattice constant). This demonstrates the importance of using experimental lattice constants (where available) for the resistivity calculations. A similar trend can also be seen for the MoNbTaVW alloy, where the residual resitivity increases with an increase in the lattice constant. 
\begin{figure}
    \centering
    \includegraphics{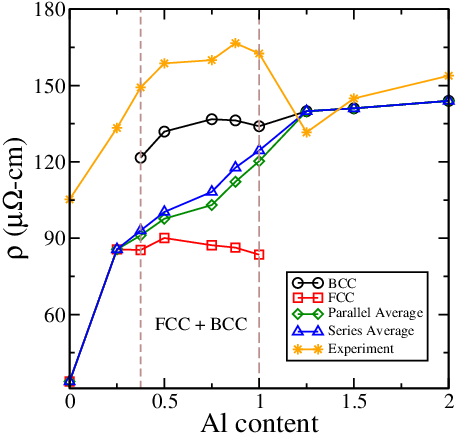}
    \caption{Residual resistivity of Al$_x$CoCrFeNi as a function of $x$, with PBE used as the exchange-correlation functional. The black curve with circles and the red curve with squares represent the residual resistivity of the pure FCC and BCC phase respectively, obtained from first-principles calculations. The blue curve with triangles and green curve with stars represents the parallel and series averages. The orange curve with crosses is the experimental result\cite{AlCoCrFeNi-1}}
    \label{fig:pbe-AlCoCrFeNi}
\end{figure}
\begin{figure}
    \centering
    \includegraphics{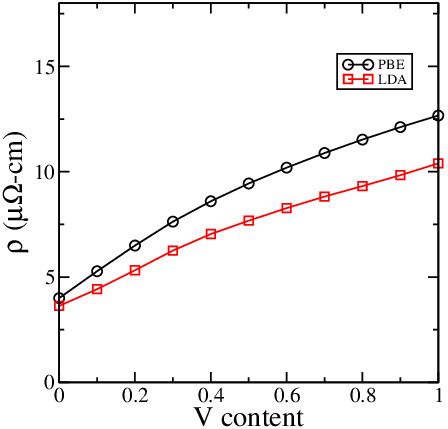}
    \caption{Residual resistivity for MoNbTaV$_x$W as a function of $x$. The black curve with circles represents calculations done using the PBE functional and the red curve with squares represents calculations done using the LDA functional.}
    \label{fig:pbe-MoNbTaVW}
\end{figure} 
\begin{figure}
    \centering
    \includegraphics{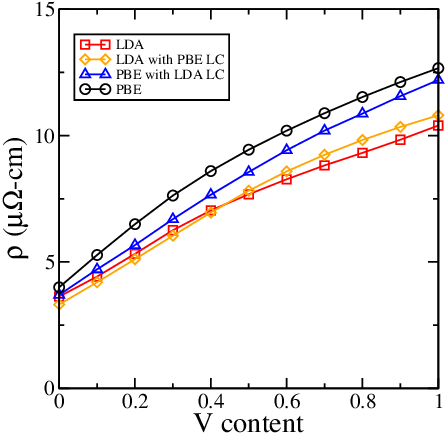}
    \caption{Residual resistivity of MoNbTaV$_x$W as a function of $x$. The black curve with circles is the LDA resistivity with LDA lattice constants, the red curve with squares is the LDA resistivity with PBE lattice constants, the orange curve with diamonds is the PBE resistivity with LDA lattice constants, and the blue curve with triangles is the PBE resistivity with PBE lattice constants.}
    \label{fig:pbe-lda-lc}
\end{figure}
\section{Exchange-Correlation Functional}
Figure \ref{fig:pbe-AlCoCrFeNi} shows the residual resistivity of Al$_x$CoCrFeNi calculated using the PBE functional. It can be seen that PBE resistivities do not match the experimental trend as well as the LDA resistivities, especially the BCC values near $x = 1$.  The sensitivity of the residual resistivity to the exchange-correlation functional was explored by Mu \textit{et al}\cite{saimu}. They find that this sensitivity is due to the variation of the exchange splitting, which controls the Fermi energy relative to the minority-spin bands. Due to LDA’s better prediction of local moments, it is a better choice for conductivity calculations in this system. Figure \ref{fig:pbe-MoNbTaVW} shows the PBE and LDA residual resistivities of MoNbTaVW. In this case, the first-principles lattice parameter for LDA and PBE are different (given in Table \ref{tab:MoNbTaVW}). To gauge the effect of lattice parameter, the PBE resistivity was calculated using LDA lattice constants and vice-versa. The results of these calculations are present in Figure \ref{fig:pbe-lda-lc}. The plot clearly shows that the effect of lattice constant is not large enough to explain the difference between the PBE and LDA resistivities. The exact reason behind the difference is unclear, but it appears that the variation is due to intrinsic differences between the two types of exchange-correlation functionals.

\begin{table}[]
\begin{tabular}{|c|c|}
\hline
$x$   & a$_{\mathrm{CuZn}}$(Atomic Units) \\ \hline
0.0 & 5.66             \\ \hline
0.05 & 5.62           \\ \hline
0.07 & 5.61           \\ \hline 
0.08 & 5.605           \\ \hline
0.09 & 5.60            \\ \hline
0.1 & 5.59             \\ \hline
0.11 & 5.585           \\ \hline
0.12 & 5.58              \\ \hline
0.13 & 5.575           \\ \hline
0.15 & 5.57            \\ \hline
0.2 & 5.55             \\ \hline
0.3 & 5.51            \\ \hline
0.4 & 5.475             \\ \hline
0.5 & 5.5445             \\ \hline
0.6 & 5.41             \\ \hline
0.7 & 5.38             \\ \hline
0.8 & 5.35             \\ \hline
0.9 & 5.33             \\ \hline
1.0 & 5.3055             \\ \hline
\end{tabular}
\label{tab:CuZn}
\caption{Lattice Constants for BCC Cu$_x$Zn$_{1 - x}$}
\end{table}
\begin{table}[]
\begin{tabular}{|c|c|}
\hline
$x$   & a$_{\rm{AgPd}}$ (Atomic Units) \\ \hline
0.0 & 7.4490\\ \hline
0.1 & 7.4796\\ \hline
0.2 & 7.5129 \\ \hline
0.3 & 7.5442    \\ \hline
0.4 & 7.5773           \\ \hline
0.5 & 7.6074           \\ \hline
0.7 & 7.6747           \\ \hline
0.8 & 7.7116           \\ \hline
0.9 & 7.7502           \\ \hline
1.0 & 7.7942           \\ \hline
\end{tabular}
\label{fig:AgPd}
\caption{Lattice Constants for FCC Ag$_x$Pd$_{1 - x}$}
\end{table}
\begin{table}[]
\begin{tabular}{|c|c|c|}
\hline
x    & a$^{\rm{LDA}}_{\rm{BCC}}$ (Atomic Units) & a$^{\rm{PBE}}_{\rm{BCC}}$ (Atomic Units) \\ \hline
0.00 & 6.00 &  6.12       \\ \hline
0.10 & 5.99 & 6.11         \\ \hline
0.20 & 5.985 & 6.105           \\ \hline
0.30 & 5.98 & 6.10            \\ \hline
0.40 & 5.97 & 6.09             \\ \hline
0.50 & 5.9601 & 6.085          \\ \hline
0.60 & 5.955 & 6.08           \\ \hline
0.70 & 5.95 & 6.07           \\ \hline
0.80 & 5.94 & 6.065            \\ \hline
0.90 & 5.935 & 6.06          \\ \hline
1.00 & 5.93 & 6.055            \\ \hline
\end{tabular}
\caption{Lattice Constants for BCC MoNbTaV$_x$W}
\label{tab:MoNbTaVW}
\end{table}